\documentclass[aps,nofootinbib,nobibnotes,noeprint,preprintnumbers,floatfix,twocolumn]{revtex4-2}
\usepackage{graphicx} % Required for inserting images
\usepackage{hyperref}
\usepackage[utf8]{inputenc}
\usepackage{graphicx}
\usepackage{amsmath}
\usepackage{xcolor}
\usepackage[caption = false]{subfig}
\usepackage{graphicx}% Include figure files
\usepackage{dcolumn}% Align table columns on decimal point
\usepackage{bm}% bold math
\usepackage{hyperref}
\usepackage{xcolor}
\usepackage{amsmath,amsfonts,amssymb}
\usepackage{commath}
\usepackage{mathtools}
\usepackage{flushend}
\usepackage[utf8]{inputenc} 
\usepackage[T1]{fontenc}

\def\tildeJ{{\widetilde{J}}}

\renewcommand{\d}{\mathrm{d}}

\newcommand{\eq}[1]{\begin{align}\begin{split}#1\end{split}\end{align}}
\begin{document}
\preprint{FERMILAB-PUB-24-0908-T}
\title{Monopole Catalyzed Baryogenesis with a $\theta$ angle}
\author{T. Daniel Brennan$^1$}
\email{tbrennan@ucsd.edu}
\author{Lian-Tao Wang$^{2,3}$}
\email{liantaow@uchicago.edu}
\author{Huangyu Xiao$^{3,4}$}
\email{huangyu@fnal.gov}
\affiliation{$^1$Department of Physics, University of California San Diego, CA 92093-0319, USA}
\affiliation{$^2$ Department of Physics \& Enrico Fermi Institute, University of Chicago }
\affiliation{$^3$Kavli Institute for Cosmological Physics, University of Chicago, Chicago, IL 60637}
\affiliation{$^4$ Theory Division, Fermilab, Batavia, IL 60510, USA}

\begin{abstract}
    Monopoles are generally expected in Grand Unified Theories (GUTs) where they can catalyze baryon decay at an unsuppressed rate by the Callan-Rubakov effect. For the first time, we show this catalysis effect can generate the observed baryon asymmetry at GeV scale temperatures. We study the minimal SU(5) GUT model and demonstrate that monopoles-fermion scattering with a $CP$-violating $\theta$-term leads to realistic baryogenesis even when $\theta\lesssim 10^{-10}$ is below the neutron EDM bound, potentially detectable in the future measurements. 
    Our calculation also shows that to generate the observed baryon asymmetry, the abundance of the monopoles is below the current experiential bounds. 
\end{abstract}

\maketitle

\section{Introduction}

Unification of gauge symmetry generally produces
monopoles in the early universe \cite{Polyakov:1974ek,tHooft:1974kcl}. In the phase transition where unified gauge symmetry is spontaneously broken, there is typically an overabundance of monopoles that is inconsistent with the current matter density, which is known as the monopole problem 
\cite{Zeldovich:1978wj,Preskill:1979zi,Albrecht:1982wi,Preskill:1984gd}. A few physical scenarios can modify the ordinary monopole relic density in phenomenologically interesting ways. 
For example, alternative cosmological scenarios, such as a second phase of inflation after the phase transition \cite{Yamamoto:1985rd,Lyth:1995ka,Harigaya:2022pjd}, can dilute the monopole relic density but not completely inflate it away. One can also have a large reheating temperature which produces monopoles from high-energy collisions of fermions, thereby enhancing the monopole density \cite{Turner:1982kh,Das:2021wei}.
These scenarios allow us to treat the monopole number density as a free parameter and study its phenomenological implications in terrestrial experiments and astrophysical observations. 

There are also various constraints placed on monopoles. 
The leading limit on the number density or the flux of magnetic monopoles comes from the Parker bound, which requires the kinetic energy gained in monopoles to not exceed the galactic magnetic energy generated by the dynamo effect \cite{Parker:1970xv,Turner:1982ag}. Monopole masses can also be bounded from below by the production of light magnetic monopoles in neutron stars  via thermal processes \cite{Gould:2017zwi}.
Direct searches of monopoles have also been conducted at colliders \cite{Bertani:1990tq,Fairbairn:2006gg,ATLAS:2012bda} (only applicable to monopole masses below TeV scale), in cosmic rays \cite{Giacomelli:2011re,Balestra:2008ps,IceCube:2015agw,IceCube:2021eye,Iguro:2021xsu,Ye:2024gsc}, and through the search for trapped magnetic monopoles in matter \cite{Price:1986ky,Ghosh:1990ki,Bendtz:2013tj}. 

In Grand Unified Theories (GUTs),  monopoles generically catalyze baryon decay at a typical strong interaction rate with an unsuppressed cross-section by the Callan-Rubakov effect \cite{Rubakov:1981rg,Callan:1982ac,Csaki:2021ozp,Brennan:2021ewu,Brennan:2023tae,vanBeest:2023mbs,Bogojevic:2024xtx,Csaki:2024ajo,Loladze:2024ayk,Hook:2024vhf,Khoze:2024hlb,Dawson:1983cm,Arafune:1983uz}. This catalysis process leads to constraints on the monopole flux from neutron stars \cite{Kolb:1982si} and white dwarfs \cite{Freese:1998es}, which are comparable to the Parker bound. It also leads to baryon decay that can be searched for by terrestrial experiments \cite{MACRO:2002iaq,Super-Kamiokande:2012tld,IceCube:2014xnp,Ahlers:2018mkf}. 

Since GUT monopoles efficiently induce baryon number violating processes, one 
might wonder if they could explain the observed baryon asymmetry in the universe. As Sakharov pointed out, this would require baryon number violation, $C$- and $CP$-violation, and departure from thermal equilibrium \cite{Sakharov:1967dj}; monopoles satisfy at least one of these conditions. However, typically the Callan-Rubakov effect leads to thermal equilibrium in monopole-fermion scattering due to the large cross-section, wiping out the baryon asymmetry in the early universe unless the monopole number density is suppressed \cite{Davis:1992ca}. 

In this letter, we present the first working example of baryogenesis via monopole-catalyzed baryon decays. We find that by explicitly breaking $CP$-symmetry via a $\theta$-angle, the Callan-Rubakov effect can be biased due to the Witten effect \cite{Witten:1979ey} so that monopoles, which are not in thermal equilibrium with the Standard Model radiation plasma, catalyze baryon and anti-baryon decays at asymmetric rates in the early universe, ultimately resulting in a net baryon asymmetry. 

In our model, baryogenesis can occur at temperatures significantly below the GUT scale, and monopole-fermion scattering can remain out of thermal equilibrium with a diluted monopole number density that avoids the overclosure of the universe.
Additionally, our scenario only requires monopole densities that are consistent with current observations and the resulting monopole relic density may provide testable signals from their magnetic charge or catalysis effect \cite{Patrizii:2015uea,Bai:2021ewf,Agrawal:2022hnf,Zhang:2024mze,Liu:2023qje,Koren:2022axd,Bai:2022nsv,Abe:2024idx}. We will show that the required monopole density is below the current experimental limit, motivating future experiments to test this minimal monopole baryogenesis scenario.

\section{Callan-Rubakov Effect with $CP$-violation}

The Callan-Rubakov effect occurs in the scattering of fermions off of a monopole. In this scattering process, the fermions probe the UV physics inside of the monopole core due to the fact that the s-wave component of the fermion wave functions are not repelled by any angular momentum or electrostatic barrier. The fermions freely interact with the UV degrees of freedom in the core, leading to a baryon number violating process whose cross-section is not suppressed by the core size. 

Consider $N$ pairs of Weyl fermions $\psi_\pm^A$ where $A=1,...,N$, with charge $\pm1$ respectively under a $U(1)_M$ gauge group.  In the presence of a Dirac monopole, the s-wave modes of the fermions are described by an effective $2d$ theory of chiral fermions $\chi,\widetilde\chi$:
\eq{
\psi^A_+=\frac{1}{r}\begin{pmatrix}
0\\\chi^A(t+r)
\end{pmatrix}\quad, \quad \psi^A_-=\frac{1}{r}\begin{pmatrix}
\widetilde\chi^A(t-r)\\0
\end{pmatrix}
}
where $\sigma^r\psi^A_\pm=\mp \psi^A_\pm$. This setup describes the low energy limit of $SU(2)_M$ gauge theory which is broken to $U(1)_M$ coupled to $N$ Weyl fermion doublets in the presence of a smooth `t Hooft-Polyakov monopole.

Note that the fermions are polarized (Zeeman effect) and are described by purely in-/out-going fermions ($\chi^A$,$\widetilde\chi^A$) according to their electric charge. Unitarity and gauge invariance of the effective $2d$ theory requires boundary conditions on the Dirac monopole core at $r=0$ relating the $\chi^A,\widetilde\chi^A$. 

The Callan-Rubakov effect is captured by a particular choice of boundary conditions at $r=0$ which can be described in terms of boundary conditions on the currents:
\eq{\label{JBC}
\left(J^A-R^A_{~B}\tildeJ^B\right)\big{|}_{r=0}=0\quad, \quad R^A_{~B}=\delta^A_{~B}-\frac{2}{N}
}
where $J^A=\bar\chi^A\chi^A$ and $\tildeJ^A=\bar{\tilde\chi}^A\tilde\chi^A$ are the in-/out-going fermion currents respectively.

If we embed the $SU(2)_M$ monopole into the Georgi-Glashow $SU(5)$ GUT with a single generation, the minimal monopole involves $4$ fermions doublets:
\begin{equation}
\begin{pmatrix}
u_2^c\\u_1
\end{pmatrix}_L,\,
   \begin{pmatrix}
u_1^c\\u_2
\end{pmatrix}_L,\,
\begin{pmatrix}
d_3^c\\e^+
\end{pmatrix}_L,\,
   \begin{pmatrix}
e^{-}\\d_3
\end{pmatrix}_L.
\end{equation}
If we more generally parametrize the $SU(2)_M$ doublets in terms of $(a^{(i)},b^{(i)})$ where $a^{(i)}=(u_2^c,u_1^c,d_3^c,e^-)_L$ and $b^{(i)}=(u_1,u_2,e^+,d_3)_L$, 
then only the $a^{(i)}$ can be incoming and only the $b^{(i)}$ can be outgoing when scattering with monopoles (their roles are reversed when scattering with anti-monopoles).  The boundary conditions \eqref{JBC} lead to the  Callan-Rubakov scattering processes:
\begin{equation}
    a^{(i)}+ a^{(j)}+ M\rightarrow \bar{b}^{(k)}+\bar{b}^{(\ell)}+M,
\end{equation}
as well as the corresponding conjugate processes 
\eq{
\bar{a}^{(i)}+ \bar{a}^{(j)}+ M\rightarrow {b}^{(k)}+{b}^{(\ell)}+M,
}
with $i\neq j \neq k \neq \ell$.
By enumerating all of these processes, we find that the the allowed baryon number violating Callan-Rubakov process for monopoles are 
\begin{equation}\label{CRMfull}
\begin{split}
 &\Delta B \qquad %q_fg_M
 \\
      & +1\qquad%\, -\qquad
      u^c_{1L}+ u^c_{2L} +M\rightarrow e^{-}_R+d_{3R}+M\\
      &-1\qquad%\, +\qquad
      u_{1R}+ u_{2R} +M\rightarrow e^{+}_L+d^c_{3L}+M \\
      & +1 \qquad%\, +\qquad 
      e^{+}_R+d^c_{3R}  +M\rightarrow  u_{1L}+ u_{2L}+M \\
       &-1\qquad%\, -\qquad 
       e^{-}_L+d_{3L}  +M\rightarrow  u^c_{1R}+ u^c_{2R}+M,
\end{split}
\end{equation}
where $\Delta B$ is the the baryon number violation in each process.

\subsection{Massive Fermions, $CP$ Violation, and the Callan-Rubakov Effect }

The theory with massive fermions can additionally depend on a $CP$-violating $\theta$-term\footnote{Note that this effect is only physical for massive fermions because the $\theta$-angle can be absorbed by a field redefinition for massless fermions.}
\eq{
S=...+\frac{i\theta}{8\pi^2}\int F\wedge F
}
where $\theta\sim \theta+2\pi$.\footnote{In the $SU(5)$ GUT theory, {a sizable and constant $\theta$ for the full $SU(5)$ is ruled out by the EDM constraints \cite{Baker:2006ts,Pendlebury:2015lrz}. The $\theta$ angle needed for our mechanism is consistent with these constraints. Alternatively, we could consider here either an effective $\theta$ in the early universe supplied by an axion-like field, or an $SU(5)$ breaking $\theta$-term that only leads to a $U(1)_Y$ $\theta$-term after GUT symmetry breaking.}} When $\theta\neq0,\pi$, this interaction violates $CP$ symmetry. Physically, the $\theta$-term endows the monopole of magnetic charge $q_m$ with an electric charge:
\eq{
q_e=-\frac{\theta}{2\pi}q_m~
}
via the Witten effect \cite{Witten:1979ey}. The $\theta$-term can be generated for example by integrating out charged fermions with a $CP$-violating mass, $m_\psi$. In this case, the electric charge is sourced by the fermion vacuum in a cloud of radius $R_c\sim1/|m_\psi|$ around the monopole \cite{Yamagishi:1982wp,Grossman:1983yf}.\footnote{Although the $\theta$-angle can be absorbed into the mass of a low energy fermion, as pointed out in \cite{Yamagishi:1982wp}, the two Hilbert spaces related by this phase rotation are not unitarily equivalent. Physically, this is because the electric charge radius of the monopole differs between these two cases. It is in this sense that we can fix the core size of the electric charge cloud of the monopole. 
} Generically, all of the charged fermions in the theory will contribute to the charge condensate near the monopole, so for simplicity we will restrict our attention to the case where the $CP$-violation comes from a $\theta$-term in the GUT theory which we can think of as activated by a massive GUT-scale fermion that results in an electric charge of fixed radius $R_c\sim 1/m_{\rm GUT}$.

The $CP$-violating $\theta$-angle also directly affects the Callan-Rubakov processes. When $\theta\neq0$, the s-wave scattering will include a potential $V(r)=\pm \theta/r$ depending on the charge of the incident fermions. Let us assume that $\theta>0$ so that the monopole has a negative charge and negatively charged fermions see a repulsive potential. 
Classically, negatively charged particles are reflected 
by the potential $V(r)=\theta/r$ with a turning point at $r_0=\theta/E$ where $E$ where $E$ is the incident energy. Quantum mechanically, they will also be totally reflected due to the fact that the potential is not integrable as $r\to 0$. 

However, when $\theta\ll 1$, the turning point for the classical particle can probe the monopole core $r_0=\theta/E\ll R_c$. In this case, we must resolve the $1/r$ singularity of the potential due to the fact that it is sourced by a finite size charge cloud. If we approximate the charge cloud as a having a constant density, then we see that an incident particle can tunnel through the repulsive electrostatic potential for $\theta/E\sim R_c$. The behavior of the scattering process is controlled by the positive parameter $\kappa=\frac{\theta}{ER_c}$: when $\kappa\ll 1$, the particle effectively scatters through the repulsive potential, whereas when $\kappa\gg1$, the particle is totally reflected. At fixed energy, dialing $\theta=0$ to $\theta>0$  decreases the efficiency of the Callan-Rubakov process for an incident negatively charged particle to zero at  $\theta\gg E R_c$.\footnote{{Turning on $\theta\neq 0$ can only decrease the efficiency of the s-wave process since they are 100\% efficient when $\theta=0$. When $\theta\neq 0$ there may also be an effect on the scattering of the higher angular momentum modes since the electrostatic potential can mitigate or enhance the angular momentum barrier. 
However, we will not consider the effect on higher angular momentum modes here. 
}
} For positively charged particles that see an attractive potential, the incident fermion will scatter off of the core of the monopole, going through the Callan-Rubakov process with 100\% efficiency.\footnote{For particles that see an  attractive potential, there are additional bound state solutions. They will not play a role in the scattering processes at the energies we consider.
}

\begin{figure}[t]\begin{center}
~\qquad\includegraphics[scale=0.7,clip, trim=0cm 0cm 8cm 0cm]{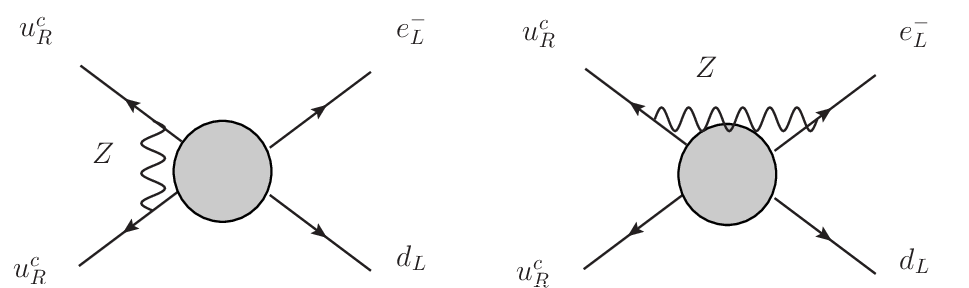}
\includegraphics[scale=0.7,clip, trim=8.5cm 0cm 0.5cm 0cm]{monopole_CP.eps}
\end{center}
\caption{The type of leading order diagrams that contributes to $A_{CP}$ in \eqref{ACPequation}. The corrections from Z bosons become relevant at temperatures  $T\sim$100 GeV.}
\label{fig:Z_boson_corrections}
\end{figure}

We will only consider scattering with $\kappa\gg1$ and $\theta\lesssim 10^{-10}$. In this case the electrostatic barrier of the monopole effectively forbids two of the Callan-Rubakov processes in \eqref{CRMfull}. Similarly, since the anti-monopole acquires a positive electric charge, the anti-monopole processes will also be suppressed so that the full set of baryon number violating Callan-Rubakov processes are effectively
\eq{
 &\Delta B \qquad %q_fg_M
 \\
      &-1\qquad%\, +\qquad
      u_{1R}+ u_{2R} +M\rightarrow e^{+}_L+d^c_{3L}+M \\
      & +1 \qquad%\, +\qquad 
      e^{+}_R+d^c_{3R}  +M\rightarrow  u_{1L}+ u_{2L}+M \\
      & +1\qquad   u^c_{1R}+ u^c_{2R} +\overline{M}\rightarrow e^{-}_L+d_{3L}+\overline{M}\\
      &-1\qquad  e^{-}_R+d_{3R}  +\overline{M}\rightarrow  u^c_{1L}+ u^c_{2L}+\overline{M}
}
Therefore, the baryon number asymmetry will be controlled by 
\eq{
\label{ACPequation}
    \Gamma( \Delta_B>0)-
      \Gamma( \Delta_B<0)\sim A_{CP} \,\Gamma_{\Delta B \neq 0},
}
where $\Gamma_{\Delta B \neq 0}$ denotes the rate of the standard Callan Rubakov process. However, since only $u^c_R, d^c_R, e^+_R$ couple to $SU(2)_L$ gauge bosons, all of the $\Delta B>0$ processes will receive corrections from weak interactions, as depicted in Fig.~\ref{fig:Z_boson_corrections}, so that $A_{CP}\sim \alpha_Z \frac{T^2}{m_Z^2}$.  At temperatures $T\sim 100$ GeV, these corrections from weak interactions to the fermion-monopole scattering become significant. 

Note that sphaleron effects occur at a temperature of the same order which can wash out the generated baryon asymmetry from the above processes. Therefore, the monopole-catalyzed baryogenesis should occur right after the electroweak phase transition.
Additionally, 
at $T\sim 100$ GeV, the Standard Model fermion condensate around the monopole has a negligible effect on the Callan-Rubakov scattering process due to the fact that the charge cloud for a given fermion $\psi$ of mass $m_\psi$ has  $\kappa_\psi\sim \theta/(R_\psi T)=\theta\frac{|m_\psi| }{T}\ll 1$ for  realistic values of $\theta$.

\section{Baryogenesis}

We will now show that if the monopole overproduction problem is resolved by some mechanism,\footnote{{For example, we can consider the scenario in which the GUT symmetry breaking, hence monopole production, happens either during the inflation or is followed by another stage of inflation. }} their scattering with Standard Model fermions is not in thermal equilibrium in the early universe so that the monopole-fermion interaction does not erase any previously generated baryon asymmetry and in fact generates baryon asymmetry when $\theta\neq 0$.

For this study, we take the cross-section of baryon-number violating interactions to be \cite{Ellis:1982bz}
\begin{equation}\label{eq:cross-section}
    \sigma_{\Delta B\neq 0}= 
    \begin{cases}
        &v^{-1}\sigma_0, \;\;\;\; T<T_0,\\
        &\sigma_0\left(\frac{T_0}{T}\right)^2, \;\; \;\; T>T_0,
    \end{cases}
\end{equation}
where $v$ is the relative velocity, $T_0\sim 1$ GeV and $\sigma_0\approx 1/T_0^2$. Note that the cross-section is not suppressed by the GUT scale. There is additionally a critical velocity, $v_{\rm c}\sim \sqrt{\alpha/4\pi} \approx 2\times 10^{-2}$, under which the cross-section is exponentially suppressed  due to interactions with the photon s-wave  \cite{Dawson:1983cm,Hook:2024vhf,Arafune:1983uz}. However, the behavior of $\sigma_{\Delta B\neq 0}$ at $v<v_{\rm c}$ will not affect baryogenesis, as it occurs in the early universe where fermion-monopole scattering is at relativistic speed, but may suppress observational signals in the late universe. 

The interaction rate for $\Delta B\neq 0$ processes can be expressed as
\begin{equation}
    \Gamma_{\Delta B\neq 0} = n_{\rm M} g_f\langle \sigma_{\Delta B\neq 0} v\rangle,
\end{equation}
where $n_{\rm M}$ is the number density of monopoles, and $g_f$ is the fermionic degrees of freedom in the thermal bath of the early universe. At high temperatures ($T>T_0$), the ratio of the interaction rate to the Hubble is
\begin{equation}
    \frac{ \Gamma_{\Delta B\neq 0}}{H}\sim 10^{-7}\Omega_{\rm M}\left(\frac{1\rm\, GeV}{T}\right)\left(\frac{10^{17}\rm\, GeV}{m_M}\right).
\end{equation}
On the other hand, the ratio is proportional to $T$ at low temperatures considering the velocity dependence of the cross-section.
Therefore, the rate of baryon number violation, that either generates or wipes-out baryon asymmetry, is maximized at $T\sim 1$ GeV. For a sufficiently  large monopole mass, this rate is never in thermal equilibrium. 

The Boltzmann equation that governs the evolution of baryon  asymmetry is
\begin{equation}
    \frac{1}{a^{3}}\frac{d( (n_B-n_{\overline{B}}) a^3)}{dt}=(\langle \sigma v\rangle_{\bar{f} M}-\langle \sigma v\rangle_{{f} M})g_fn_fn_{\rm M},
\end{equation}
where we have approximated $n_f\approx n_{\bar{f}}$. 
The baryon asymmetry is then produced from the out-of-thermal equilibrium scattering between monopoles and fermions.
The baryon asymmetry can be defined as the baryon-to-entropy ratio
\begin{equation}
    Y = \frac{n_B-n_{\overline{B}}}{\frac{2\pi^2}{45}g_{*S}(T)T^3},
\end{equation}
which should be matched with the experimental value $Y=8.718\times 10^{-11}$ where $g_{*S}$ is the effective number of degrees of freedom.
The Boltzmann equation can be rewritten as
\begin{equation}
    \frac{dY}{d{\rm ln}a} = \frac{\Delta \sigma \,g_f n_f }{H}\,\frac{n_{\rm M}}{\frac{2\pi^2}{45}g_{*S}(T)T^3},
\end{equation}
where $\Delta \sigma \approx A_{CP} \,\langle \sigma v\rangle_{f M} $ characterizes the difference in cross-sections  due to $CP$-violation,   
$n_f = \frac{3\zeta(3)}{2\pi^2}T^3$ is the number density of quarks in equilibrium at high temperature, and $\frac{n_{\rm M}}{\frac{2\pi^2}{45}g_{*S}(T)T^3}$ is a constant, which can be related to the monopole relic density in the Current Universe as
\begin{equation}
   \Omega_{\rm M} = \frac{ n_{\rm M}m_M}{\rho_{\rm crit}} =\frac{n_{\rm M}}{\frac{2\pi^2}{45}g_{*S}(T_{})T_{}^3} \frac{\frac{2\pi^2}{45}g_{*S}(T_{\rm CMB})T_{\rm CMB}^3 m_M}{\rho_{\rm crit}}.
\end{equation}
Since $A_{CP}$ is large at $T\sim 100$ GeV, we can solve for the baryon asymmetry produced at that temperature and take $g_f=80$. Note that the baryon asymmetry could be further enhanced if $A_{CP}$ is also large at $T\sim 1$ GeV, which represents the optimal temperature for monopole-catalyzed baryogenesis. However, in this work, we proceed with the less optimal temperature.

Solving the Boltzmann equations, the baryon asymmetry generated from monopole-catalyzed decays is
\begin{equation}
    Y \simeq 8.718\times 10^{-11} \left( \frac{m_M}{10^{17}\,\rm GeV}\right)^{-1} \left(\frac{A_{CP}  \,\Omega_{\rm M}}{10^{-2}}\right)
\end{equation}
Therefore we only need $ A_{CP} \,\Omega_{\rm M} \approx  10^{-2}$ for successful baryogenesis if monopole mass is $m_M = 10^{17}$ GeV. This is below the current experimental bound for sizable $CP$-violation. 

The predicted monopole flux responsible for baryogenesis is given by
\begin{equation}
\label{fluxfinal}
    \Phi_{\rm M} = \frac{n_{\rm M}\, v_{\rm M}}{4\pi} = 5 \times 10^{-15} v_{\rm M} {\,\rm cm^{-2}s^{-1}sr^{-1}},
\end{equation}
where $v_{\rm M}$ is the velocity of monopoles in units of speed of light. 
Note that this result is independent of the monopole mass, as the generation of baryon asymmetry only depends on the number density of monopoles. But if the magnetic monopoles are lighter than $10^{16}$ GeV, they are accelerated by the galactic magnetic field and their velocity can be approximated as $v_{\rm M} \approx 3 \times 10^{-3} ({10^{16} \, \rm GeV}/{m_M})^{1/2}$ \cite{Turner:1982ag} (capped by the speed of light). Hence, the Parker bound provides a lower bound on the monopole mass, $m_M\gtrsim 10^{12}$ GeV.

From \eqref{fluxfinal}, we see that the monopole flux is still consistent with the Parker bound ($\Phi \lesssim 10^{-15} {\,\rm cm^{-2}s^{-1}sr^{-1}} $) if monopoles have a speed of  $v_{\rm M} \lesssim 0.2$.\footnote{The extended Parker bound, obtained by considering the survival
of an early seed field \cite{Adams:1993fj,Lewis:1999zm}, is potentially more constraining for relativistic monopoles, but may be alleviated by the fact that monopoles are accelerated by intergalactic magnetic fields \cite{Perri:2023ncd}. } The monopole flux $\Phi_{\rm M}$  is also bounded from the catalysis of nucleon decays in white dwarfs, which requires $\Phi_{\rm M} < 1.9 \times 10^{-14} \, v_{\rm M} ^2 \, {\rm cm^{-2} s^{-1} sr^{-1}}$ for catalysis at the strong interaction rate \cite{Freese:1998es}. However, the proposed exponential suppression of the Callan-Rubakov effect for $v_{\rm M} \lesssim v_{\rm c}\sim 2\times 10^{-2}$ \cite{Dawson:1983cm,Hook:2024vhf,Arafune:1983uz} suggests that 
the bound from the catalysis effect on white dwarfs is not applicable for low-velocity monopoles ($m_M\gtrsim 10^{14}$ GeV). For the same reason, the catalysis effect of GUT monopoles, if not accelerated to velocities above the threshold, may not be present in terrestrial experiments, including MACRO \cite{MACRO:2002iaq}, Super-Kamiokande \cite{Super-Kamiokande:2012tld}, and IceCube \cite{IceCube:2014xnp}. Therefore, the required monopole number density to generate the observed baryon asymmetry is consistent with all the experimental and observational constraints given the preferred mass range for GUT monopoles. 

\section{Conclusion}
In this paper we studied fermion-monopole scattering in theories with a $CP$-violating $\theta$-term in the early universe and demonstrate how it can produce baryon asymmetry with a non-thermal population of GUT monopoles. In this scenario, the Sakharov conditions are naturally satisfied in a universe with GUT monopoles: a non-thermal monopole relic density at a temperature much lower than the GUT scale creates a departure from thermal equilibrium, baryon-number violation occurs in monopole-fermion scattering via the Callan-Rubakov effect at an unsuppressed rate, and $C$ and $CP$ are violated by weak interactions and the $\theta$-term.

As a concrete example, we explicitly show the baryon-number-violating scattering in the Georgi-Glashow SU(5) GUT has a preference for generating more matter than anti-matter with a positive $\theta$-angle in the GUT theory. In this case, the $CP$-violation critically endows monopoles with an electric charge from the Witten effect, creating Coulomb potential that favors certain $\Delta B\neq 0$ processes. The weak interactions then enhance the monopole-fermion scattering for $\Delta B>0$ processes which become significant at temperatures $T\gtrsim 100$ GeV. These ensure an asymmetric production rate of matter versus antimatter, even when monopoles and anti-monopoles are present in equal numbers. For this scenario we only require that $\theta\gg T/m_{\rm GUT}\sim 10^{-14}$ which is satisfied for the range $10^{-10}>\theta\gg 10^{-14}$ and is consistent for experimental bounds on the neutron EDM \cite{Baker:2006ts,Pendlebury:2015lrz}, and potentially detectable in future measurements.  In this scenario, the predicted monopole density required for baryogenesis, if not depleted by other mechanisms, is still consistent with current experimental constraints, providing additional motivation for the ongoing search for magnetic monopoles.

\section*{Acknowledgement}
We thank Paddy Fox, Seth Koren, Csaba Csaki, Ken Van Tilburg, Simon Knapen, Peter Graham, Aneesh Manohar, Gordan Krnjaic, and Savas Dimopoulos for their helpful discussions. HX is supported by Fermi Research Alliance, LLC under Contract DE-AC02-07CH11359 with the U.S. Department of Energy. TDB is supported by Simons Foundation award 568420 (Simons Investigator) and award 888994 (The Simons Collaboration on Global Categorical Symmetries). LTW is supported by 
the Department of Energy grant DE-SC0013642.

\bibliographystyle{apsrev4-2}
\bibliography{ref}

\end{document}